\documentclass[12pt]{article}

\usepackage{amssymb,latexsym,amsmath}

\usepackage{fullpage}
\usepackage{nicefrac}
\usepackage{mathrsfs}%different \cal using \mathscr
\usepackage{slashed}% for dslash

\usepackage{graphicx}
\usepackage{epsfig}

\makeatletter \@addtoreset{equation}{section}
\makeatother

\newcommand{\be}{\begin{equation}}
\newcommand{\ee}{\end{equation}}
\newcommand{\bea}{\begin{eqnarray}}
\newcommand{\eea}{\end{eqnarray}}
\newcommand{\nn}{\nonumber}

\begin{document}
%%%%%%%%%%%%%%%%%%%%%%%%%%%%%%%%%%%%%%%%%%%%%%%%%%%%%%%%%%%%%%

\begin{titlepage}
	\thispagestyle{empty}
	\begin{flushright}
		\hfill{SISSA 21/2011/EP}
	\end{flushright}
	
	\vspace{90pt}
	
	\begin{center}
	    { \LARGE{\bf Lifshitz black holes in string theory}}\\ [3mm]

			\vspace{30pt}
		
		{Irene Amado$^{\,1}$ and Anton F. Faedo$^{\,2}$}
		
		\vspace{25pt}
		
		{\small
		{\it ${}^1$  SISSA \\
		Via Bonomea 265, 34136 Trieste, Italy}
		
		\vspace{15pt}
		
		{\it ${}^2$  INFN, Sezione di Padova \\
		Via Marzolo 8, 35131 Padova, Italy}
		
		\vspace{15pt}

		{iamado @ sissa.it , faedo @ pd.infn.it}
		}
		
		\vspace{90pt}

		{\bf Abstract}
	\end{center}
We provide the first black hole solutions with Lifshitz asymptotics found in string theory. These are expected to be dual to models enjoying anisotropic scale invariance with dynamical exponent $z=2$ at finite temperature. We employ a consistent truncation of type IIB supergravity to four dimensions with an arbitrary 5-dimensional Einstein manifold times a circle as internal geometry. New interesting features are found that significantly differ from previous results in phenomenological models. In particular, small black holes are shown to be thermodynamically unstable, analogously to the usual AdS-Schwarzschild black holes, and extremality is never reached. This signals a possible Hawking--Page like phase transition at low temperatures.

\vspace{10pt}

\end{titlepage}

\baselineskip 6 mm

%%%%%%%%%%%%%%%%%%%%%%%%%%%%%%%%%%%%%%%%%%%%%%%%%%%%%%%%%%%%%%
%%%%%%%%%%%%%%%%%%%%%%%%%%%%%%%%%%%%%%%%%%%%%%%%%%%%%%%%%%%%%%
\newpage

%%%%%%%%%%%%%%%%%%%%%%%%%%%%%%%%%%%%%%%%%%%%
\section{Introduction}
%%%%%%%%%%%%%%%%%%%%%%%%%%%%%%%%%%%%%%%%%%%%

The targets of the gauge/gravity duality have been vastly increased with the appreciation that it can be applied to strongly coupled condensed matter systems \cite{Hartnoll, Herzog, McGreevy}. Among the phenomena of interest we encounter phase transitions governed by fixed points enjoying the anisotropic scaling  $t\to\lambda^z \,t$ while $x\to\lambda\, x$. The parameter $z$ is termed dynamical exponent, and $z=1$ corresponds to the familiar scale invariance of the conformal group. Clearly, a departure from this value breaks conformality and thus the gravity dual cannot be an AdS space. 

Two different holographic realizations of this symmetry have been proposed. The first one is expected to model cold atoms at unitarity and the corresponding metrics were constructed in \cite{Son, Balasubramanian}. For the particular value $z=2$ the isometry group of this space coincides with the invariance group of the Schr\"odinger equation and in consequence they are called Schr\"odinger geometries.  

The second avatar, Lifshitz geometries, is the focus of this paper. This class of metrics are named after the Lifshitz field theory that is supposed to arise as the long wavelength (continuous) limit of lattice models of strongly correlated electrons. They were introduced in \cite{Kachru} precisely as gravity duals of such systems and read
\begin{equation}\label{LifMet}
ds^2\,=\,L^2\,\left(-r^{2z}\,dt^2+\,r^2\,dx^\ell dx^\ell+\,\frac{dr^2}{r^2}\right)\,,\quad\quad\ell=1,\dots,d-2
\end{equation}
where $L$ sets the scale of the spacetime and for $z=1$ one recovers AdS in Poincar\'e coordinates. In addition to anisotropic scaling if $r\to\lambda^{-1}\,r$, the isometries of these metrics include spacetime translations along with space rotations. 

Often the approach to condensed matter holography is phenomenological, with no manifest connection to string or M-theory. The matter content of the gravitational model is selected with the sole purpose of supporting the metrics of interest and give rise to the desired properties. For instance, the Lifshitz metric (\ref{LifMet}) is a solution of AdS gravity coupled to a massive vector \cite{Taylor} if one tunes the mass and the cosmological constant to $m^2=z\,(d-2)\,L^{-2}$ and $2\Lambda=-\left[z^2+(d-3)z+(d-2)^2\right]\,L^{-2}$ respectively. Notice that both the dynamical exponent and the dimensionality of the spacetime enter in these expressions.

The flexibility of the phenomenological models is certainly an advantage and much insight can be gained with this modus operandi. Nonetheless, gauge/gravity dualities are firmly established just in the context of string theories and one could be skeptical about the legitimacy of bottom-up constructions. In other words, it is important to verify that the relevant solutions can also be obtained within string or M-theory. Unfortunately, as one can presume, this is a more intricate task. 

However, many important steps have been taken in this direction. By now we have at our disposal embeddings into string theory of holographic superconductors \cite{Gubser, Gauntlett} as well as the Schr\"odinger spaces mentioned above \cite{Maldacena, Adams, Herzog2}. Lifshitz geometries have proven to be more elusive, but they were recently found in \cite{Balasubramanian2, Donos, Gregory, Donos2, Lif_from_AdS}.

Consistent truncations have been instrumental in obtaining various of these solutions. They provide lower-dimensional simplified setups with respect to 10 or 11-dimensional supergravities. Compared to phenomenological models they have the virtue that the embedding of the solution into string theory is automatic. The drawback is rigidity, in the sense that the parameters of the theory, such as masses and charges, are fixed during the reduction. Coming back to the example of Lifshitz this means that, either one is working with a consistent truncation containing a massive vector and a cosmological constant with the exact values given above, or the solution is not there. One could then rephrase the search for Lifshitz solutions in string theory as the chase for a vector with the correct mass.

An important point was the observation that one can keep the appropriate massive modes in the reduction of higher-dimensional supergravities \cite{Maldacena}. Stimulated by this result, several consistent truncations containing massive vectors were subsequently found \cite{Gauntlett2, Cassani, Liu, Gauntlett3, Cassani2, Bena, Donos2}. Nevertheless, none of the vectors included in those truncations seemed to have the correct mass (except one in \cite{Donos2} that gives $z\sim39$). A proper vector to support Lifshitz metrics with $z=2$ was identified in \cite{Balasubramanian2} and further exploited to construct infinite families of solutions in \cite{Donos, Lif_from_AdS}.

On the other hand, it is known since the dawn of the AdS/CFT duality that introducing temperature in the gauge theory corresponds to placing a black hole in the bulk of the gravity side. When the gauge theory is conformal, the corresponding black hole has AdS asymptotics in the boundary. Similarly, since the Lifshitz field theory is endowed with anisotropic scale invariance, the pertinent black holes describing its finite temperature generalization must have the metric (\ref{LifMet}) as asymptotic geometry. 

Black holes possessing this property have been extensively studied in phenomenological models, starting with the $d=4$, $z=2$ numerical ones of \cite{Danielsson}. Extensions to other dimensionalities and critical exponents are available as well \cite{Bertoldi, Giribet, Giribet2, Bertoldi2}. The charged case is also interesting for condensed matter applications \cite{LifSuper, Pang, Mann5}. Lifshitz black holes are also solutions of higher-order theories of gravity \cite{Cai, Mann2, Mann3, Mann4} and Brans--Dicke models \cite{Giribet3}. Even if most of the results are numeric, a few exact black holes are known \cite{Mann, Balasubramanian3}. Thermodynamic properties of these systems are detailed in \cite{Bertoldi3, Cheng}. 

Despite all this considerable effort, a proper embedding of Lifshitz black holes into string theory was still missing\footnote{The solutions of \cite{Hartong} are rather naked singularities than genuine black holes.}. In this work, we construct an infinite family of $d=4$, $z=2$ Lifshitz black holes in type IIB supergravity. Building on the results of \cite{Donos, Lif_from_AdS}, we present a consistent truncation based on ${\rm E}_5\times {\rm S}^1$, where E$_5$ is an arbitrary 5-dimensional Einstein space, that contains a massive vector suitable for supporting the solution. In addition it incorporates two scalars, the dilaton and the modulus of the circle, whose presence complicates the study of the equations but triggers interesting new features with respect to the more minimal phenomenological models. In particular, we find no extremal limit and an instability for small black holes similar to the usual AdS-Schwarzschild one. This signals to a possible Hawking--Page like transition \cite{Hawking} in these systems. 

The structure of the paper is simple: in section \ref{model} we construct the consistent truncation and present the 4-dimensional model and its equations of motion, in section \ref{numerics} we solve numerically the equations and comment on the properties of the black holes found and in section \ref{conclusions} we conclude.

%%%%%%%%%%%%%%%%%%%%%%%%%%%%%%%%%%%%%%%%%%%%
\section{The model}\label{model}
%%%%%%%%%%%%%%%%%%%%%%%%%%%%%%%%%%%%%%%%%%%%

In accordance with the results of \cite{Lif_from_AdS}, a 4-dimensional Lifshitz space with $z=2$ can be attained starting from a model in five dimensions containing 
an axion field and admitting an AdS$_5$ vacuum. The mechanism is the following: if the 5d theory is reduced on ${\rm S}^1$ while using the axion field strength to thread a 
flux, the vector gauging the circle isometry has the correct mass to support the 4d Lifshitz solution. In this section we will provide a very simple model, coming from
 a consistent truncation of type IIB supergravity, that meets these requirements and thus is suitable for the embedding of Lifshitz-like solutions into string theory. Using an ansatz that generalizes the metric (\ref{LifMet}) to allow for a horizon, we derive the complete set of equations to be solved.

A straightforward way to get an AdS$_5$ solution in type IIB is to compactify on a 5-dimensional Einstein space E$_5$ with a Freund--Rubin ansatz for the RR 5-form. 
On the other hand, the most natural candidate for the axion would be the RR scalar $C_0$. Indeed, as already mentioned in \cite{Lif_from_AdS}, the model obtained in
 this manner is a consistent truncation of type IIB supergravity as long as one retains in addition the dilaton $\phi$. In detail, we take the 10-dimensional metric 
to be
\be
ds^2\,=\,ds^2({\rm M}_5)+ds^2({\rm E}_5)\,,
\ee
where M$_5$ is for the time being an arbitrary 5-dimensional spacetime and the internal Einstein metric is normalized as $R_{ij}=4\,g_{ij}$. 
By assumption the axio-dilaton has sole dependence on the M$_5$ coordinates and from the set of forms the only non-vanishing one is
\be
F_5\,=\,4\,\left(1+*_{10}\right)\,{\rm Vol}({\rm E}_5)\,,
\ee
ensuring self duality. We denoted ${\rm Vol}( {\rm E}_5)$ the volume form on the Einstein space. Within this ansatz the Bianchi identities are satisfied 
(taking $F_1=dC_0$), as well as all the 10-dimensional equations of motion except the one for the axio-dilaton and the external components of Einstein's equations, that 
follow from the 5d action
\begin{eqnarray}
 S&=&\frac{1}{2\kappa_5^2}\,\,\int\,\left[R*1 -\, \frac{1}{2}d\phi\wedge*d\phi \,-\, \frac{1}{2}\,e^{2\phi}\,dC_0\wedge*dC_0\,+\,12\,*1\right]\,.
\end{eqnarray}
Notice the presence of a negative cosmological constant $\Lambda=-6$ that emerges from a combination of the internal space curvature and the 5-form flux. 
It is then apparent that the model contains an AdS$_5$ vacuum for constant scalars. In order to find the 4-dimensional Lifshitz solution one needs to further reduce 
the theory on a circle with metric
\be
ds^2({\rm M}_5)\,=\,e^{-T}\,ds^2({\rm M}_4)+\,e^{2T}\,\left(d\vartheta+\mathcal A\right)\otimes\left(d\vartheta+\mathcal A\right)\,.
\ee
The scalar $T(x)$ is a modulus parameterizing the size of the S$^1$ while the vector $\mathcal A(x)$ is gauging the reparameterization invariance of the coordinate $\vartheta$ on the circle. Both depend exclusively on the coordinates of the 4d spacetime, denoted collectively $x$. This vector turns out to support a Lifshitz metric if one St\"uckelberg couples it to the axion via a flux on the S$^1$. In doing so, the correct ansatz for the axion is $C_0(x,\vartheta)\,=\,C_0(x)\,+\,n\,\vartheta$ with $n$ a constant. We should point out 
that, even if $C_0$ has a explicit linear dependence on the circle coordinate, its axionic nature ensures that it necessarily enters into the action covered by a derivative and
 thus the circle reduction is consistent. Moreover, in this way one generates a covariant derivative for it and consequently a mass term for the vector
\be
dC_0(x,\vartheta)\,=\,dC_0(x)\,+\,n\,d\vartheta=\,DC_0(x)\,+\,n\,\left(d\vartheta+\mathcal A\right)\,,
\ee
with  $DC_0=dC_0-n\,\mathcal A$. The second term will enter as a modification to the potential. Assuming that the dilaton depends just on the M$_4$ coordinates the 
reduction is then straightforward and the resulting action reads
\begin{eqnarray}
 S&=&\frac{1}{2\kappa_4^2}\,\,\int\,\Bigg[R*1-\frac32 dT\wedge*dT -\, \frac{1}{2}d\phi\wedge*d\phi \,-\, \frac{1}{2}e^{2\phi}\,DC_0\wedge*DC_0\nonumber\\[2mm]
&&\qquad\qquad-\frac12\,e^{3T}d\mathcal A\wedge*d\mathcal A+\left(12\,e^{-T}-\frac{n^2}{2}\,e^{2\phi-3T}\right)\,*1\,\Bigg]\,,
\end{eqnarray}
where the gravitational constant has been redefined to $\kappa_5^2=\kappa_4^2\int_{{\rm S}^1}d\vartheta$. For our purposes it is more convenient to dualize the axion to a 
2-form and work with a system more similar to the one in \cite{Kachru}. Let us add to the action the following piece
\begin{equation}
 S'\,=\,\frac{1}{2\kappa_4^2}\,\,\int\,\left(G_1+n\,\mathcal A\right)\wedge dA_2\,,
\end{equation}
where we denoted $G_1=DC_0$. Variation with respect to $A_2$ gives the Bianchi $dG_1=-n\,\mathcal F$ as expected, while varying w.r.t. $G_1$ one deduces instead the 
duality condition $G_1=e^{-2\phi}\,*dA_2$ that once substituted into the action yields
\begin{eqnarray}
 S&=&\frac{1}{2\kappa_4^2}\,\,\int\,\Bigg[R*1-\frac32 dT\wedge*dT -\, \frac{1}{2}d\phi\wedge*d\phi \,-\, \frac{1}{2}e^{-2\phi}\,dA_2\wedge*dA_2\nonumber\\[2mm]
&&\qquad\qquad-\frac12\,e^{3T}\mathcal F\wedge*\mathcal F+\left(12\,e^{-T}-\frac{n^2}{2}\,e^{2\phi-3T}\right)\,*1\,+n\,A_2\wedge\mathcal F\Bigg]\,.
\end{eqnarray}
Leaving apart the scalars $\phi$ and $T$, this is the form of the action used in \cite{Kachru}. These two modes are spurious for the purposes of obtaining Lifshitz 
solutions, but are unavoidable for consistency of the truncation. Indeed, the scalar $T$ will be present in any model that uses $\mathcal A$ as a support for the 
Lifshitz metric, since it is well known that the modulus of a circle cannot be consistently switched off while retaining the vector gauging its isometry. Similarly, 
we cannot truncate away the dilaton and simultaneously keep the axion. The presence of these spectator fields complicates the analysis of the equations of motion with respect to the phenomenological models of \cite{Kachru, Danielsson}. On the other hand, some new appealing features of our black holes will be caused by these modes, as we will detail.

We want to generalize the Lifshitz metric (\ref{LifMet}) above to allow for black hole solutions with $z=2$, so we pick\footnote{Here we set the size of the space to $L=1$ for simplicity. All the dimensionful quantities in the following are measured in units of $L$.}
\begin{equation}\label{metric}
ds^2\,=\,-r^4\,f^2(r)\,dt^2+r^2\,d\Omega_2^2+\frac{g^2(r)}{r^2}\,dr^2\,,
\end{equation}
where the transverse piece reads
\be
d\Omega_2^2\,=\,\left\lbrace
  \begin{array}{lcl}
     d\psi^2+\sin^2{\psi}\,d\chi^2 &\qquad\text {if}\qquad &k=1\\[2mm]
     d\psi^2+\psi^2\,d\chi^2&\qquad\text {if}\qquad &k=0 \\[2mm]
     d\psi^2+\sinh^2{\psi}\,d\chi^2&\qquad\text {if}\qquad &k=-1
  \end{array}
  \right.
\ee
and $k$ is the curvature corresponding to spherical, flat or hyperbolic horizon, respectively. Since we are interested in geometries that are asymptotically Lifshitz we will impose the boundary conditions $f(r),\ g(r)\to1$ as $r\to\infty$. 

To support this metric we take the following ansatz for the forms (with the obvious choice of vielbeins)
\begin{eqnarray}
\mathcal F&=&2\,\,h(r)\,\,\theta^r\wedge\theta^t\,,\nonumber\\[3mm]
F_3&=&\frac{4}{n}\,\,j(r)\,\,\theta^r\wedge\theta^\psi\wedge\theta^\chi
\end{eqnarray}
and permit a radial dependence of the scalars
\begin{equation}
T\,=\,T(r)\hspace{2cm}{\rm and}\hspace{2cm}e^{-2\phi}\,=\,\frac{n^2}{4}\,e^{-2\varphi(r)}\,.
\end{equation}
Substituting this ansatz into the matter e.o.m. we get the following system of coupled differential equations
\begin{eqnarray}
&&\left[\,e^{3T}\,h\,r^2\right]'-2\,j\,g\,r\,=\,0\,,\label{eoms}\\[3mm]
&&\left[\,e^{-2\varphi}\,j\,f\,r^2\right]'-2\,h\,f\,g\,r\,=\,0\,,\label{eomj}\\[3mm]
&&\left[\frac{\varphi'\,r^5\,f}{g}\right]'\,\frac{1}{f\,g\,r^3}+4\,e^{-2\varphi}\,j^2-4\,e^{2\varphi-3T}\,=\,0\,,\label{eomphi}\\[3mm]
&&\left[\frac{T'\,r^5\,f}{g}\right]'\,\frac{1}{f\,g\,r^3}+2\,e^{3T}\,h^2-4\,e^{-T}+2\,e^{2\varphi-3T}\,=\,0\,,
\end{eqnarray}
where $'$ indicates derivative with respect to the radial coordinate. There are two combinations of Ricci tensor components, namely $R_{tt}+R_{rr}\pm 2\,R_{\psi\psi}$, that give first order equations once substituted into Einstein's equations. They read
\begin{equation}\label{eomf}
\frac{2\,r\,f'}{f}\,=\,-5+\frac34\,(T'\,r)^2+\frac14\,(\varphi'\,r)^2+g^2\left(e^{-2\varphi}\,j^2-e^{3T}\,h^2+6\,e^{-T}-e^{2\varphi-3T}+\frac{k}{r^2}\right)
\end{equation}
and
\begin{equation}\label{eomg}
\frac{2\,r\,g'}{g}\,=\,3+\frac34\,(T'\,r)^2+\frac14\,(\varphi'\,r)^2+g^2\left(e^{-2\varphi}\,j^2+e^{3T}\,h^2-6\,e^{-T}+e^{2\varphi-3T}-\frac{k}{r^2}\right)\,.
\end{equation}
The third component is not independent and follows from these two.

One important point is that, since $f'\propto f$, we can use (\ref{eomf}) in the e.o.m. of the matter fields to decouple $f$ from the system. We can then solve 
for $T$, $\varphi$, $g$, $h$ and $j$ and substitute in the equation for $f$. We have not found exact black hole solutions of the complete system (\ref{eoms})-(\ref{eomg})\footnote{Plugging the metric and form functions of the topological black hole in \cite{Mann} one can not verify the equation of the dilaton (\ref{eomphi}).}.

It is simple to check that the Lifshitz metric (\ref{LifMet}) with $z=2$ and $d=4$ verifies the equations above. This amounts to take 
\begin{equation}\label{LifSol}
f\,=\,g\,=\,h\,=\,j\,=1\qquad\qquad{\rm and}\qquad\qquad \varphi\,=\,T\,=0\,.
\end{equation}
Therefore, the solutions we are interested in will approach these values asymptotically in the boundary located at $r\to\infty$.

%%%%%%%%%%%%%%%%%%%%%%%%%%%%%%%%%%%%%%%%%%%%%%%%%%%%%%%%%%%%%%%
\section{The Lifshitz black hole solutions}\label{numerics}
%%%%%%%%%%%%%%%%%%%%%%%%%%%%%%%%%%%%%%%%%%%%%%%%%%%%%%%%%%%%%%%

In order to find asymptotically Lifshitz black hole solutions we will resort to numerics. But before integrating the system, let us study the asymptotic behaviour of such solutions.

\subsection{Asymptotic behaviour}

In the UV regime, $r\to\infty$, we are looking for solutions that asymptote to Lifshitz geometry with scaling $z=2$, i.e. $f,\,g,\,h,\,j\to 1$ and $\varphi,\,T\to 0$ near the boundary. Therefore we can linearize the equations of motion around the fixed point and study the subleading behaviour of the fields in the asymptotically large $r$ region. For convenience let us define new scalar fields
\be
\tau(r)=e^{T(r)}\quad ;\quad\gamma(r)= e^{\varphi(r)}\,,
\ee
that approach $\tau, \gamma\to 1$ for the Lifshitz solution. If we expand the fields around the scale invariant solution as $\psi(r)=\psi_{\rm \small{Lif}} + \delta \psi(r)$, it is straightforward to solve analytically the resulting equations of motion up to linear order in $\delta\psi$. The most important feature of the linearized system is the presence of a zero mode solution, i.e. independent of the radial coordinate. Of course, coming back to the non-linearized problem, such zero mode develops an $r$-dependent profile. If one gets close enough to the boundary, the zero mode gives the dominant subleading contribution to the fields, since one expects the other modes to have already decayed. However, even if the zero mode has negative amplitude, its decay will be very slow and so will be the convergence to Lifshitz geometry\footnote{ It is easy to check that our non-linearized system is solved by a series expansion in a marginally decaying mode of the form $c(r)=\log{(\log{r})}/\log{r}$.}. For that reason we will seek solutions for which the amplitude of the zero mode vanishes and the decaying modes of the fields behave as $1/r^2$ and $\log{r}/r^2$. In particular, the leading orders of the solutions to the non-linearized system in the UV, imposing Lifshitz asymptotics and a vanishing amplitude for the zero mode, are given by 
\bea
f(r)&=&1-\frac{3 \beta+\alpha-k}{2 r^2} - \frac{8 \alpha-k}{4 r^2} \log{r} +\dots\,, \nonumber\\
g(r)&=&1-\frac{6 \beta-2 \alpha-k}{4 r^2} + \frac{8 \alpha-k}{4 r^2} \log{r} +\dots\,, \nonumber\\
h(r)&=&1-\frac{\alpha}{r^2} +\dots\,, \\
j(r)&=&1-\frac{3 \beta-5 \alpha}{2 r^2} - \frac{8 \alpha-k}{4 r^2} \log{r} +\dots\,, \nonumber\\
\gamma(r)&=&1+\frac{\alpha}{r^2} +\dots\,, \nonumber\\
\tau(r)&=&1+\frac{\beta}{r^2} + \frac{8 \alpha - k}{6 r^2} \log{r} +\dots\,,\nonumber
\eea
where $\alpha$ and $\beta$ are arbitrary constants and the dots represent terms that decay faster than $1/r^2$.

In the deep IR we want the above solutions to flow to a black hole like geometry. A non-extremal black hole is characterized by the presence of a non-degenerate horizon at a finite $r=r_{\rm H}$. In our coordinates, this implies that the $g_{tt}$ component of the metric has a simple zero at this point, whereas the $g_{rr}$ component has a simple pole there. Regularity of the solution further demands that the 2-form field strength $j(r)$ vanishes at the horizon as $\sim 1/g(r)$. We will assume that the vector field strength and the scalar fields have a finite value at the horizon. Under these considerations, the near-horizon expansion of the fields takes the form
\begin{eqnarray}
f(r)&=& \sqrt{r-r_{\rm H}} \left(f_0 + f_1 (r-r_{\rm H}) + f_2 (r-r_{\rm H})^2 + \dots\right)\,,\nn\\[2mm]
g(r)&=& \frac{1}{\sqrt{r-r_{\rm H}}} \left(g_0 + g_1 (r-r_{\rm H}) + g_2 (r-r_{\rm H})^2 + \dots\right)\,,\nn\\[2mm]
j(r)&=& \sqrt{r-r_{\rm H}} \left(j_0 + j_1 (r-r_{\rm H}) + j_2 (r-r_{\rm H})^2 + \dots\right)\,,\nn\\[2mm]
\label{irexp}h(r)&=& h_0 + h_1 (r-r_{\rm H}) + h_2 (r-r_{\rm H})^2 + \dots\,,\\[2mm]
\gamma(r)&=& \gamma_0 + \gamma_1 (r-r_{\rm H}) + \gamma_2 (r-r_{\rm H})^2 + \dots\,,\nn\\[2mm]
\tau(r)&=& \tau_0 + \tau_1 (r-r_{\rm H}) + \tau_2 (r-r_{\rm H})^2 + \dots\,.\nn
\end{eqnarray}
Plugging that expansion into the equations of motion and solving order by order we find that $f_0$ just appears as a global factor in the $f(r)$ field. This is because its equation of motion is linear in $f(r)$ so the normalization of the field is not fixed. This fact will be important for the numerical integration. Once we rescale $f(r)$ by $f_0$, all the coefficients can be written only in terms of $\,h_0,\,\gamma_0$ and $\tau_0$, being the first ones in the series given by
\begin{eqnarray}
g_0&=&\frac{\sqrt{r_{\rm H}}}{\sqrt{\frac{k}{r_{\rm H}^2}+\frac{6}{\tau_0}-\frac{\gamma_0^2}{\tau_0^3}-h_0^2 \tau_0^3}}\,,\\
j_0&=&\frac{2 h_0 \gamma_0^2}{\sqrt{r_{\rm H}}\sqrt{\frac{k}{r_{\rm H}^2}+\frac{6}{\tau_0}-\frac{\gamma_0^2}{\tau_0^3}-h_0^2 \tau_0^3}}\,,\\
&\vdots & \nonumber
\end{eqnarray}
It is clear that the horizon values of the vector field strength and the scalars are constrained by demanding the solutions to be real. For the planar and spherically symmetric black hole solutions\footnote{We will not further consider the hyperbolic case $k=-1$.} it implies the following upper bounds
\begin{equation}\label{bounds}
|h_0| < \frac{1}{\tau_0^2} \sqrt{\frac{k \,\tau_0}{r_{\rm H}^2}+6-\frac{\gamma_0^2}{\tau_0^2}}\qquad{\rm and}\qquad \frac{\gamma_0}{\tau_0}<\sqrt{\frac{k \,\tau_0}{r_{\rm H}^2}+6}\,,
\end{equation}
where $\gamma_0$ and $\tau_0$ are both positive and from now on we assume $h_0$ to be so.

\subsection{Numerical integration}

Now we can proceed to the numerical integration of the system. We use the near-horizon expansions in (\ref{irexp}) to generate initial data  close to the horizon and then integrate the equations of motion in (\ref{eoms})-(\ref{eomg}) towards the UV demanding that the solution has Lifshitz behaviour at the boundary. Since we have a large number of free parameters the strategy is the following. As we already know, $f_0$ is nothing but a normalization scale for $f(r)$. Therefore we can set $f_0=1$ and afterwards rescale with the asymptotic boundary value of the numerical solution, automatically fixing the normalization of $f$ in such a way that $f(r)\to 1$ as $r\to \infty$. We are then left with four parameters, i. e. the horizon radius, the vector field strength and the scalar horizon values. In each integration we keep fixed the black hole size $r_{\rm H}$ and the modulus $\tau_0$, while tuning the dilaton $\gamma_0$ and the 1-form flux $h_0$ in order to get Lifshitz solutions. The precision of our numerical results is $|\psi-1|<10^{-4}$ at the cutoff $r=10^6$, where $\psi$ represents all the fields but $f$. We then explore the parameter space by repeating the procedure for different pairs of initial data $\left\lbrace r_{\rm H},\tau_0\right\rbrace$. In principle we could have picked any other free parameter rather than $\tau_0$ to generate the initial data. However this is a natural choice since the modulus $T$ will be present in any Lifshitz solution based on the massive vector $\mathcal{A}$ \cite{Donos, Lif_from_AdS}.

\begin{figure}[!htbp]
\centering{
\includegraphics[scale=0.9]{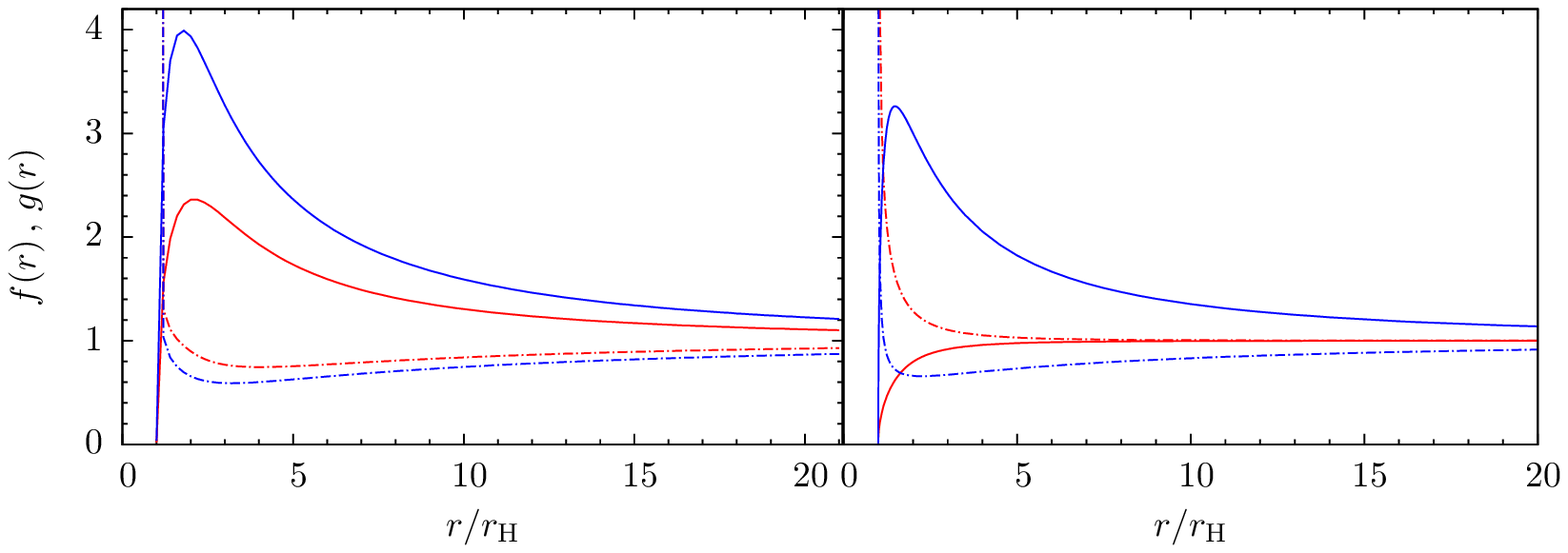}
}
\caption{\label{plots1} Metric functions $f(r)$ (solid) and $g(r)$ (dashdotted) for various different pairs of initial values. The red (inner) and the blue (outer) lines correspond to $\tau_0=2$ and $\tau_0=1$ for $r_{\rm H}=1/2$ (left), and to $\tau_0=2$ and $\tau_0=1/2$ for $r_{\rm H}=20$ (right), respectively.}
\end{figure}

\begin{figure}[!htbp]
\centering{
\includegraphics[scale=0.9]{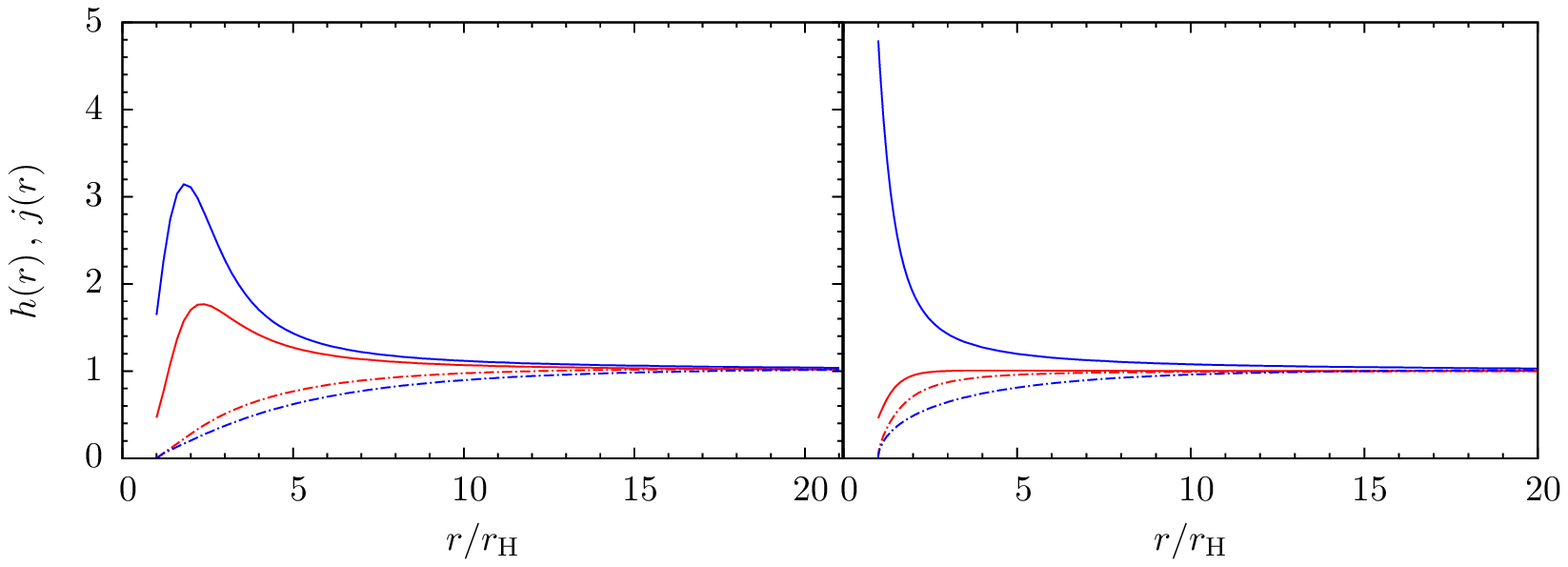}
}
\caption{\label{plots2} Form fields $h(r)$ (solid) and $j(r)$ (dashdotted) for various different pairs of initial values. The red (inner) and the blue (outer) lines correspond to $\tau_0=2$ and $\tau_0=1$ for $r_{\rm H}=1/2$ (left), and to $\tau_0=2$ and $\tau_0=1/2$ for $r_{\rm H}=20$ (right), respectively.}
\end{figure}

\begin{figure}[!htbp]
\centering{
\includegraphics[scale=0.9]{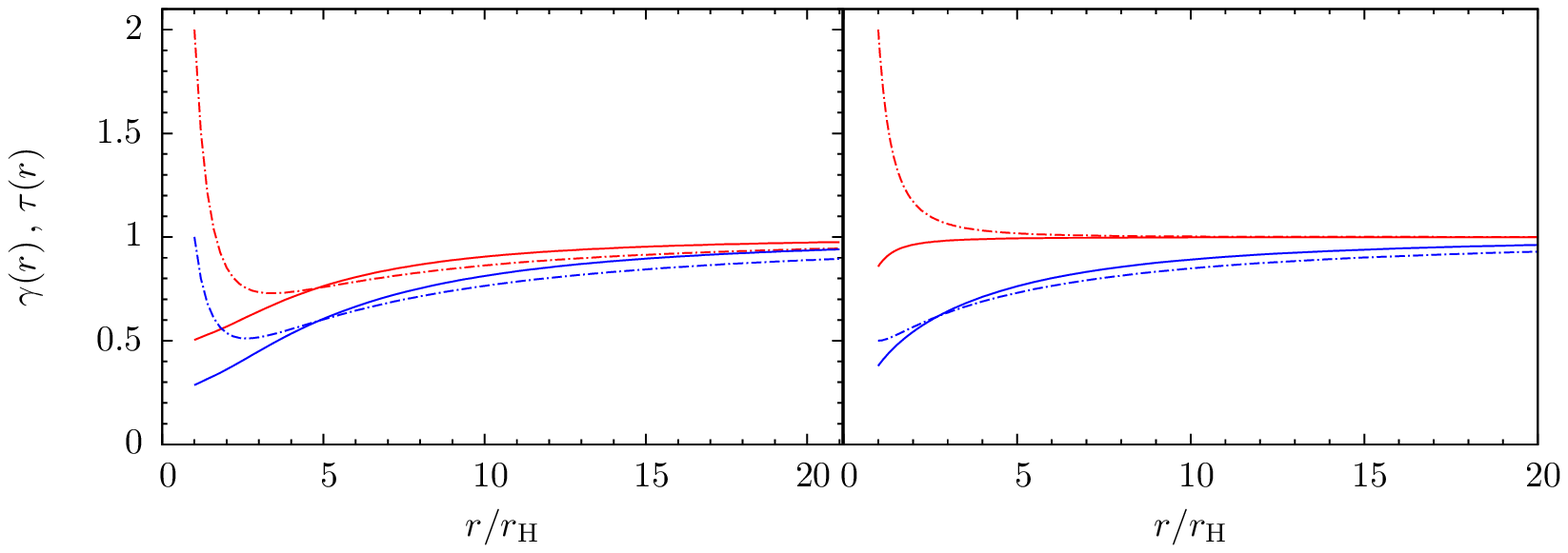}
}
\caption{\label{plots3} Scalar fields $\gamma(r)$ (solid) and $\tau(r)$ (dashdotted) for various different pairs of initial values. The red (upper) and the blue (lower) lines correspond to $\tau_0=2$ and $\tau_0=1$ for $r_{\rm H}=1/2$ (left), and to $\tau_0=2$ and $\tau_0=1/2$ for $r_{\rm H}=20$ (right), respectively.}
\end{figure}

\begin{figure}[!htbp]
\centering{
\includegraphics[scale=0.8]{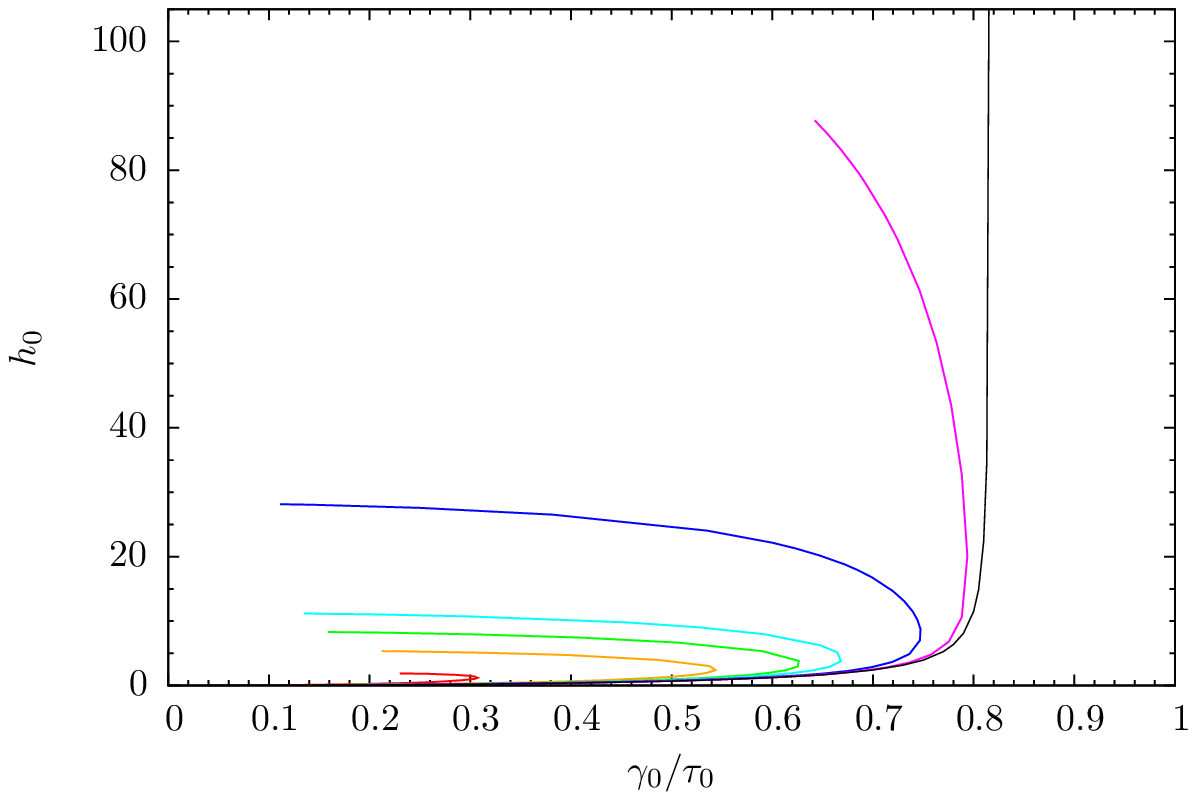}
}
\caption{\label{plots4} Horizon 1-form flux $h_0$ as a function of the ratio of the scalar fields $\gamma_0/\tau_0$ for fixed $r_{\rm H}=1/2,\,1,\,3/2,\,2,\,5,\,20$ and $r_{\rm H}=\infty$, from left to right.}
\end{figure}

In figures \ref{plots1}, \ref{plots2} and \ref{plots3} are shown the metric, form and scalar radial profiles of the solutions for two different black hole sizes and each for two different values of the modulus. It is clear that convergence to Lifshitz geometry is better for large values of the scalar $\tau_0$ at the horizon. Convergence is also better the larger the black hole for a given $\tau_0$. It turns out that $\gamma(r)\leq1$ for any pair of initial values of the parameters. This implies that the dilaton in the black hole solution is always smaller than in the pure Lifshitz space, i. e. $\phi_{\rm BH}(r)\leq\phi_{\rm Lif}$. On the other hand, the constraint found on the modulus is simply given by $\tau_0>\gamma_0>0$ (see figure \ref{plots4}), so the original modulus $T(r)$ can take both positive and negative values in the black hole solution. The peak and dip developed by the metric functions of the black holes close to the horizon grow when the modulus $\tau_0$ decreases. 

From this behaviour and the fact that convergence is worse the smaller is $\tau_0$, one could expect that for a given black hole size, there is a minimum value of that scalar field at the horizon for which the field solutions are able to stabilize to their Lifshitz limit when flowing to the UV. In fact this is the case. In figure \ref{plots4} it is plotted the 1-form flux as a function of $\gamma_0/\tau_0$ for fixed values of the horizon radius\footnote{In principle, for a given horizon radius, $h_0$ is a function of both $\gamma_0$ and $\tau_0$ and not just of its ratio. However, in the $k=0$ case the system is invariant under rescaling of the radial coordinate and solutions are independent of $r_{\rm H}$. Furthermore, solutions are related by rescaling of the fields in such a way that the space of free parameters is reduced to $\lbrace h_0,\,\gamma_0/\tau_0 \rbrace$. On the other hand, the flat case $k=0$ is equivalent to the $r_{\rm H}\to\infty$ limit of the spherically symmetric black hole. Comparison of the flat and spherical black hole solutions is possible by plotting $h_0$ vs. $\gamma_0/\tau_0$.}. Very large values of $\tau_0$ correspond to the beginning of the curves, where $h_0\to0$, whereas the ending of the curves corresponds to the minimum value of $\tau_0$ for each $r_{\rm H}$. At this minimum, $\gamma_0$ is also a minimum, while $h_0$ reaches its maximum value. For the flat black hole solution, $k=0$, the limiting configuration corresponds to the maximum value of the ratio of the scalars, $\gamma_0/\tau_0=0.8164$, and to the minimum value that the scalars can reach. 

The bounds (\ref{bounds}) are satisfied along the whole curves. Moreover, the bound for $h_0$ is saturated only in the $\tau_0\to\infty$ limit, where $h_0\to0$. For very large values of the modulus $\tau_0$, the $\gamma$ field is roughly constant, $\gamma(r)\lesssim1$, so there is no running of the dilaton with the RG flow, and the horizon 1-form flux almost vanishes. Therefore the black hole is only supported by the modulus $T$. Notice that this behaviour is independent of the black hole size and that for non-vanishing $r_{\rm H}$ extremality is only achieved in the limit $\tau_0\to\infty$, that of course is not a regular solution at the horizon. For regular solutions extremal black holes could appear in the limit of vanishing black hole size. This kind of zero temperature solutions were found in phenomenological models \cite{Danielsson,Mann}. However, it will not be the case in the stringy model at hand due to the presence of the additional scalars.

\begin{figure}[!tbp]
\centering{
\includegraphics[scale=0.9]{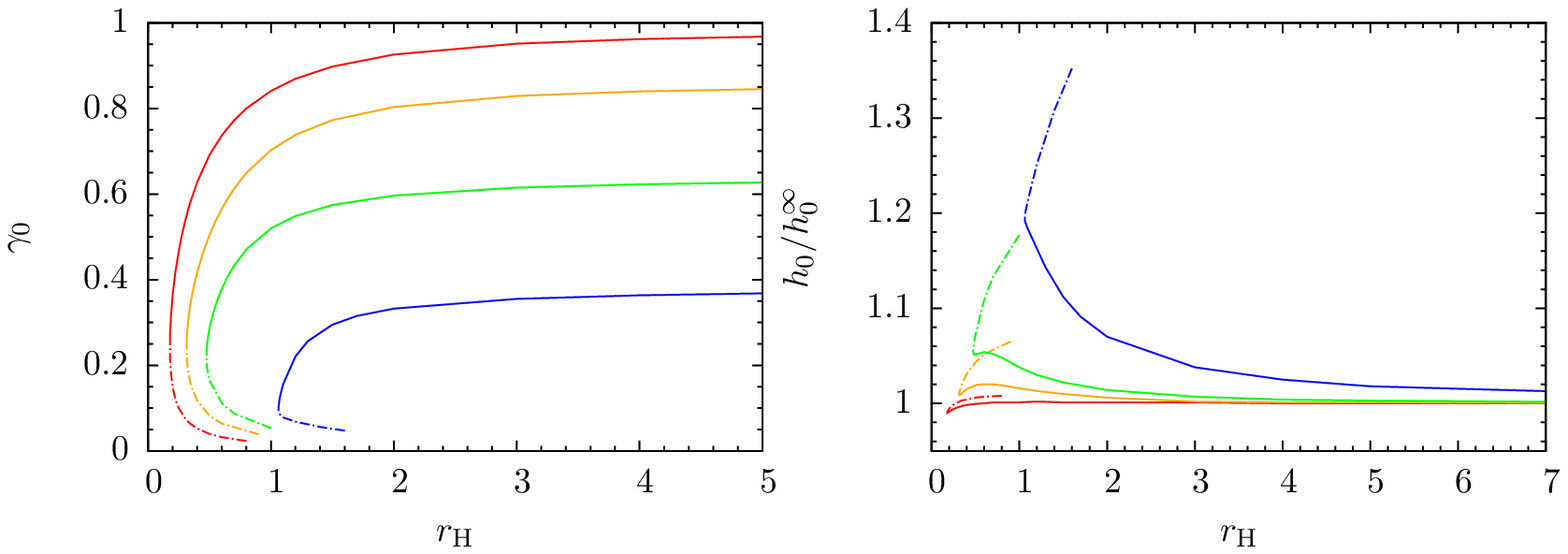}
}
\caption{\label{plots5} (Left) Horizon dilaton $\gamma_0$ as a function of $r_{\rm H}$ for fixed $\tau_0=6,\,2,\,1,\,1/2$, from top to down. (Right) Horizon 1-form flux $h_0$ as a function of $r_{\rm H}$ for fixed $\tau_0=6,\,2,\,1,\,1/2$, from down to top, normalized by the flat solution value: $h_0^{\infty}=0.0549,\,0.459,\,1.57,\,4.79$, respectively. First branch (solid) is connected to the $k=0$ case, while the second branch (dashdotted) is unstable.}
\end{figure}

For a given initial $\tau_0$ there is a minimum black hole size that can support the configuration of fluxes that make the solution flow to Lifshitz asymptotics. In figure \ref{plots5} we show the horizon values $\gamma_0$ and $h_0$ as a function of $r_{\rm H}$ for fixed values of $\tau_0$. For large $r_{\rm H}$ they take the asymptotic values given by the flat solution, $k=0$, and both of them remain barely constant in a wide range of $r_{\rm H}$. When the horizon radius gets close to its minimum size, the dilaton $\gamma_0$ rapidly decreases to its minimum value, while the flux $h_0$ either grows or decreases depending on the value of $\tau_0$. There is a second branch of solutions that is not connected to the flat solution and that merges smoothly with the one described above at the minimum $r_{\rm H}$. Over this second branch, $\gamma_0$ decreases and $h_0$ increases while the black hole grows. In principle, we expect $\gamma_0$ to asymptotically vanish for large black hole radius. Convergence to Lifshitz geometry of these solutions is very poor when one moves further from the minimum size, contrasting with the first branch, where convergence is better the larger the black hole. The second branch will be shown to be thermodynamically unstable. 

\subsection{Thermodynamics of Lifshitz black holes}

From the ansatz (\ref{metric}) and using the near-horizon expansion (\ref{irexp}), we can determine various thermodynamical quantities that characterize the black holes. The Hawking temperature and entropy density are given by
\bea
\label{temp}T_{\rm H}&=&\frac{r_{\rm H}^3 f_0}{4\pi g_0} \,,\\[2mm]
\label{entr}s&=&\pi r_{\rm H}^2\,,
\eea
both for $k=0,1$, after appropriate volume normalization and where we set the Newton constant to one. The conditions (\ref{bounds}) have to be satisfied in order to have a positive real temperature. In fact, if the bound for $h_0$ is saturated the black hole becomes extremal. Due to the intrinsic dependence of the horizon values of the fields on the black hole size it is not possible to write the explicit dependence of the entropy density on the temperature. However approximate relations can be found in certain regions of the parameter space.

\begin{figure}[!tbp]
\centering{
\includegraphics[scale=0.9]{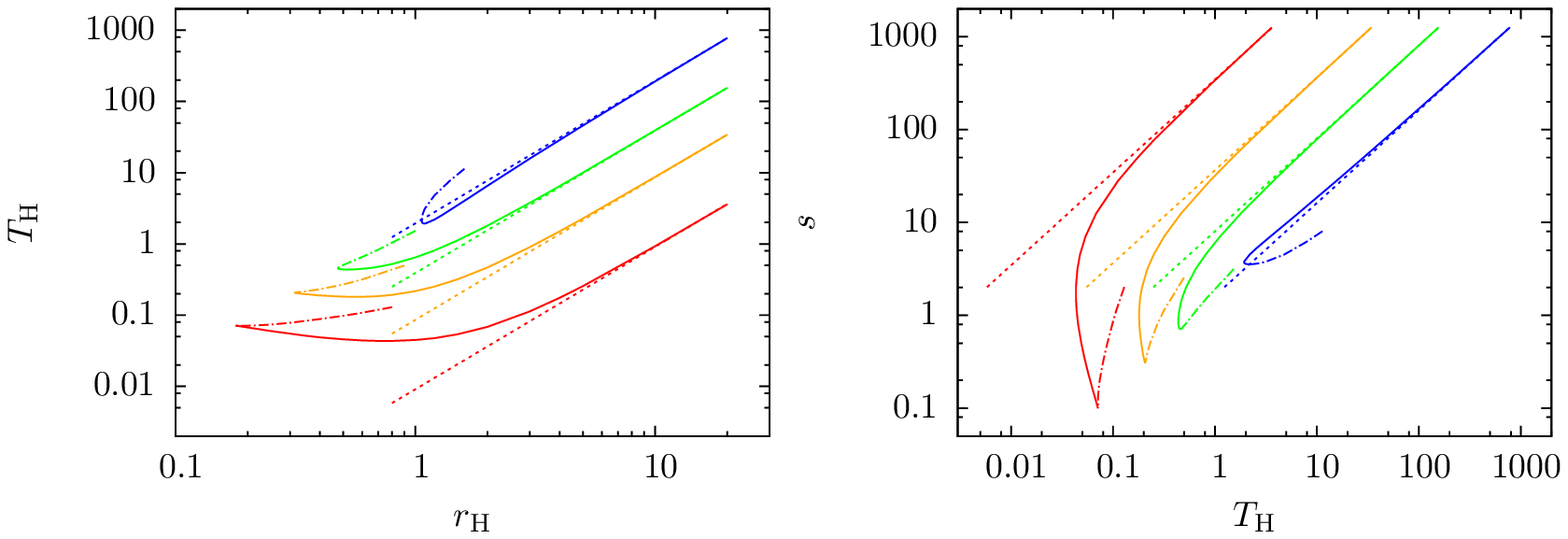}
}
\caption{\label{plots6} (Left) Temperature as a function of $r_{\rm H}$ and (right) entropy density as a function of temperature of the first (solid) and second (dashdotted) branches for fixed $\tau_0=6,\,2,\,1,\,1/2$, from down to top (left) and from left to right (right). It is also shown the asymptotic behaviour for large values of radius (dotted).}
\end{figure}

The thermodynamic behaviour of the Lifshitz black holes above can be easily studied. Figure \ref{plots6} shows the Hawking temperature as a function of the black hole size and the entropy density as a function of the temperature. There are significant differences with previous phenomenological models. As already mentioned, in \cite{Danielsson,Mann} extremal black holes were found in the limit of vanishing horizon radius since temperature for such models turned out to be a monotonic function of the black hole size. In our string theory model, the presence of additional matter fields prevents the horizon to shrink to $r_{\rm H}=0$ and consequently extremality is never reached. Temperature for large black holes increases with the size, while for small black holes it develops a negative slope. Having negative specific heat, these small black holes are always thermodynamically disfavoured. Notice that the concept of smallness is $\tau_0$-dependent. This is precisely the behaviour shown by AdS-Schwarzschild black holes. In analogy with this well known case \cite{Hawking}, we expect to have a Hawking--Page like phase transition between large Lifshitz black holes at high temperature and `thermal' Lifshitz (pure Lifshitz space with compact Euclidean time) at low temperature.

For very large black holes the temperature is approximately $T_{\rm H}\propto r_{\rm H}^2$. Since the Bekenstein--Hawking entropy density is simply given by the area of the horizon, it is possible to write down its temperature dependence in the $r_{\rm H}\gg1$ limit as 
\be 
s=\alpha_{\tau_0}\, T_{\rm H}\,,
\ee
where the proportionality constant can be computed numerically. For the solutions presented in figure \ref{plots6}, it reads
\be
\alpha_{6}=345.2\,,\quad\alpha_{2}=36.53\,,\quad\alpha_{1}=8.055\,,\quad\alpha_{1/2}=1.619\,.
\ee 
Hence, large Lifshitz black holes satisfy the expected scaling behaviour for a field theory with dynamical exponent $z=2$ in flat $2+1$-dimensions, $s\propto T^{2/z}$.

\begin{figure}[!tbp]
\centering{
\includegraphics[scale=0.9]{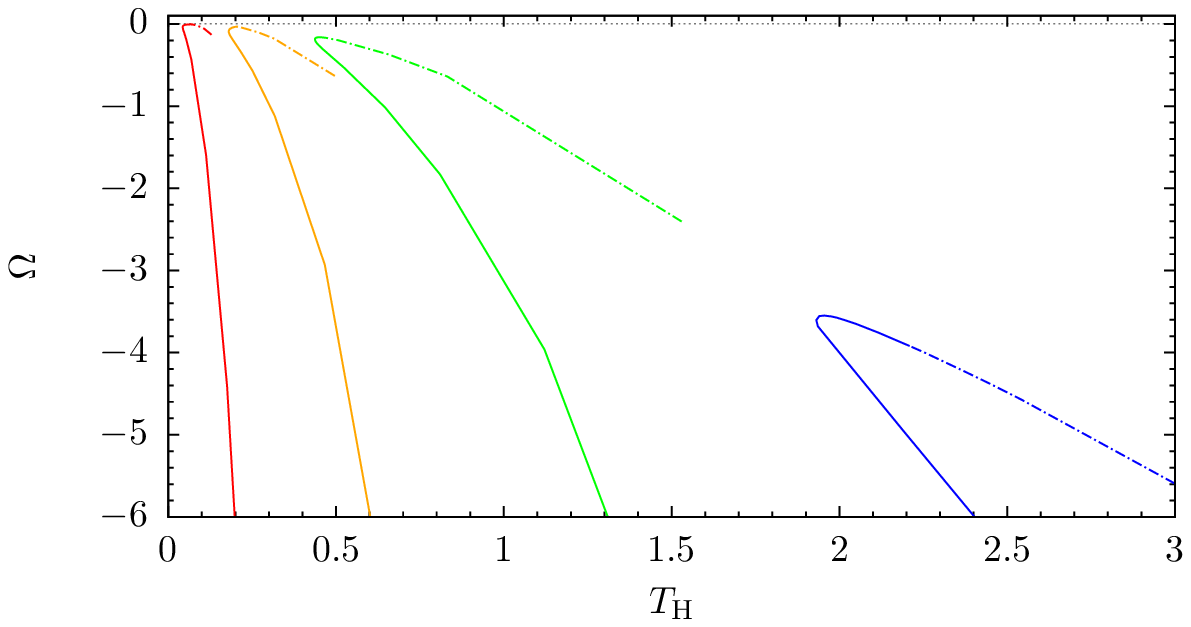}
}
\caption{\label{plots7} Free energy density as a function of temperature of the first (solid) and second (dashdotted)  branches for fixed $\tau_0=6,\,2,\,1,\,1/2$, from left to right.}
\end{figure}

In order to ensure that the small black holes and the second branch of solutions are unstable compared to the large ones we can compute the free energy of the system. From the first law of thermodynamics, $\epsilon + P = s T$, and the relation between energy density and pressure for the dual field theory, $\epsilon = P$, the free energy is easily computed as
\be 
\Omega = -P = -\frac{1}{2} s T\,.
\ee
This thermodynamic relation has been proved to hold by computing the free energy from the Euclidean bulk action in \cite{Bertoldi2,Balasubramanian3,Cheng,Ross} for various phenomenological models\footnote{In our case we have extra scalar fields not considered in the mentioned works. However it is not expected that the presence of additional scalars spoils this result once the appropriate counterterms are taken into account.}. Figure \ref{plots7} shows the free energy as a function of the temperature for various initial values of $\tau_0$. For temperatures above the minimum temperature, there exist two different solutions. The lower one, hence the stable, corresponds to large black holes, whereas the upper one is clearly unstable and corresponds both to small black holes and to the second branch of solutions, as expected from previous considerations. For temperatures below the minimum temperature, Lifshitz black holes can not exist and presumably a phase transition to `thermal' Lifshitz will take place. 

%%%%%%%%%%%%%%%%%%%%%%%%%%%%%%%%%%%%%%%%%%%%%%%%%%%%%%%%%%%%%%%
\section{Conclusions}\label{conclusions}
%%%%%%%%%%%%%%%%%%%%%%%%%%%%%%%%%%%%%%%%%%%%%%%%%%%%%%%%%%%%%%%

In this work we have constructed infinite families of asymptotically Lifshitz black holes in string theory. We made use of a consistent truncation of type IIB supergravity on an arbitrary 5-dimensional Einstein space that keeps the axio-dilaton. By further reducing on a circle we obtain an additional scalar (the modulus of the S$^1$) and a vector that is appropriate to support Lifshitz black holes with dynamical exponent $z=2$. The model obtained in this way is not minimal, in the sense that previous phenomenological studies were lacking the dilaton and the modulus that are mandatory for the consistency of our truncation. 

These extra modes contribute in a very non-trivial way to the equations and yield new interesting features. We found no extremal black holes. In phenomenological models \cite{Danielsson, Mann} the extremal limit corresponds to vanishing black hole horizon, but here we observe a lower bound on the size. On top of that, small black holes turn out to be unstable. For large black holes, temperature increases with the horizon radius, while for small black holes it develops a negative slope, pointing to an instability that we confirmed by computing the free energy. The fact that black holes have a minimum temperature leads us to expect a Hawking--Page like phase transition, resembling the AdS-Schwarzschild case.

It would be interesting to know if this behaviour is general in string theory constructions. It may happen that this property is common to solutions based on the vector we kept, that comes always accompanied by the modulus of the circle. This could be checked by using truncations with matter content different from the axio-dilaton, for instance the one based on the T$^{1,1}$ of \cite{Donos}. Other suitable truncations should be easy to find \cite{Lif_from_AdS}. 

A pressing question is the nature of the expected phase transition. The first step would be to compute the free energy of the thermal Lifshitz space, in order to proof its existence. We must point out that for Lifshitz geometries the gauge/gravity holographic dictionary is underdeveloped, though some progress has been made \cite{Costa}. This makes involved the identification of the relevant degrees of freedom in each of the phases. In consequence, it might be premature to interpret the phase transition as the analogue of confinement-deconfinement \cite{Witten} in non-relativistic systems.  We hope to come back to these issues in the near future. 

It is certainly possible to extend our results to charged Lifshitz black holes. When the internal manifold is not only Einstein but admits a Sasakian structure, there is a consistent truncation that includes an additional vector. Switching on this vector introduces a charge density and thus charged black hole solutions similar to Reissner--Nordstr\"om can be obtained. Interestingly, the truncation can be enhanced to include a charged scalar \cite{Gubser, Cassani, Lif_from_AdS}. Lifshitz black hole solutions where this scalar condenses and breaks the Abelian symmetry would be dual to the superfluid phase of a non-relativistic superconductor \cite{LifSuper}.

%%%%%%%%%%%%%%%%%%%%%%%%%%%%%%%%%%%%%%%%%%%%%%%%%%%%%%%%%%%%%%%
\bigskip
\section*{Acknowledgments}

\noindent We are grateful to Davide Cassani and Gianguido Dall'Agata for many insightful discussions, careful reading of the manuscript and valuable comments. The work of A.~F. is partially supported by the Padova University project CPDA105015/10.

%%%%%%%%%%%%%%%%%%%%%%%%%%%%%%%%%%%%%%%%%%%%%%%%%%%%%%%%%%%%%%%

%%%%%%%%%%%%%%%%%%%%%%%%%%%%%%%%%%%%%%%%%%%%%%%%%%%%%%%%%%%%%%
  
%%%%%%%%%%%%%%%%%%%%%%%%%%%%%%%%%%%%%%%%%%%%%%%%%%%%%%%%%%%%%%%
\end{document}